\documentclass[superscriptaddress,aps,prl,floats,twocolumn,twoside,floatfix]{revtex4}
\usepackage{epsfig}
\usepackage{amsmath}
\usepackage{amssymb}
\newcommand{\av}[1]{\mathbb E\left(#1\right)}

\newcommand{\be}{\begin{equation}}
\newcommand{\ee}{\end{equation}}
\newcommand{\bea}{\begin{eqnarray}}
\newcommand{\eea}{\end{eqnarray}}
\newcommand{\s}{\sigma}

\begin{document}

\title{Overlap Equivalence in the Edwards-Anderson model}  

\author{Pierluigi Contucci}
\affiliation{
Universit\`{a} di Bologna, Piazza di Porta S.Donato 5, 40127 Bologna, Italy}

\author{Cristian Giardin\`a}
\affiliation{Eurandom, P.O. Box 513 - 5600 MB Eindhoven, The Netherlands}

\author{Claudio Giberti}
\affiliation{
Universit\`a di Modena e Reggio Emilia, via Fogliani 1, 42100 Reggio Emilia, Italy}

\author{Cecilia Vernia}
\affiliation{
Universit\`{a} di Modena e Reggio Emilia, via Campi 213/B, 41100 Modena, Italy}

\begin{abstract}
We study the relative
fluctuations of the link overlap and the square standard overlap in the three dimensional
Gaussian Edwards-Anderson model with zero external field. We first analyze the correlation
coefficient and find that the two quantities are uncorrelated above the critical
temperature. Below the critical temperature we find that the link overlap has vanishing
fluctuations for fixed values of the square standard overlap and large volumes.
Our data show that the conditional variance scales to zero in the thermodynamic limit.
This implies that, if one of the two random variables tends to a
trivial one (i.e. delta-like distributed), then also the other does
and, by consequence, the TNT picture should be dismissed. We
identify the functional relation among the two variables using the
method of the least squares which turns out to be a monotonically
increasing function. Our results show that the two overlaps are
completely equivalent in the description of the low temperature
phase of the Edwards-Anderson model.
\end{abstract}

\maketitle

The low temperature phase of short-range spin-glasses is among the most unsettled problems
in condensed matter physics \cite{MPV,NS}. To detect its nature it was originally proposed an order
parameter by Edwards and Anderson \cite{EA}, the disorder average of the local squared magnetization
\begin{equation}
q_{EA} = {\rm Av}(\omega_i^2) = {\rm Av}\left(\left[\frac{\sum_{\sigma}\sigma_i e^{-\beta H_{\sigma}}}
{\sum_{\sigma}e^{-\beta H_{\sigma}}}\right]^2\right),
\end{equation}
which coincides with the quenched expectation of the local {\em standard} overlap of two
spin configurations drawn according to two copies of the equilibrium state carrying identical disorder
\begin{equation}
{\rm Av}(\omega_i^2) = <q_i> = {\rm Av}\left(\frac{\sum_{\sigma,\tau}\sigma_i \tau_i \;
e^{-\beta (H_{\sigma}+H_{\tau})}}{\sum_{\sigma,\tau}e^{-\beta
(H_{\sigma}+H_{\tau})}}\right).
\end{equation}
The previous parameter should reveal the presence of frozen spins in random directions at
low temperatures. While that choice of the local observable is quite natural, it is far to be
unique; one can consider, for instance, the two point function ${\rm Av}(\omega_{ij}^2)$.
In the case of nearest neighbour correlation function this yields the quenched average of the
local {\em link} overlap.

When summed over the whole volume, link overlap and standard overlap give rise to a priori
different global order parameters. In the mean field case the two have a very simple relation:
in the Sherrington-Kirkpatrick (SK) model, for instance, it turns out that the link overlap coincides
with the square power of the standard overlap up to thermodynamically irrelevant terms.
But in general, especially in the finite dimensional case of nearest neighbor interaction
like the Edwards-Anderson (EA) model, the two previous quantities have a different behavior with
respect to spin flips: when summed over regions the first undergoes changes of volume sizes
after spin-flips, while the second is affected only by surface terms.

From the mathematical point of view, their role is also quite different. The square of the
standard overlap represents, in fact, the covariance of the Hamiltonian function for the SK model,
while the link overlap is the covariance for the EA model. Two different overlap definitions
are naturally related to two different notions of distance among
spin configurations. It is an interesting question to establish if two distances are equivalent
for the equilibrium measure in the large volume limit and, if yes, to what extent
(see \cite{Pa} for a broad discussion on {\em overlap equivalence} and its relation with ultrametricity).
They could in fact be simply equivalent in preserving neighborhoods (topological equivalence)
or they could preserve order among distances (metric equivalence).
The a-priori different properties of the two overlaps have also been discussed
in relation to the different pictures (droplet \cite{FH}, mean-field \cite{MPV}, TNT \cite{KM,PY}),
that have been proposed
to describe the nature of the low temperature spin-glass state.
In this respect, the distributions of the two overlaps are expected to be delta-like 
(trivial distribution, droplet theory), to have support on a finite interval 
(non trivial distribution, mean-field theory), or to have different behaviour
depending on which overlap is considered (trivial link overlap distribution, 
non trivial standard overlap distribution, TNT theory).

In this paper we consider the EA model in d=3, with Gaussian couplings and zero external
magnetic field in periodic boundary conditions. We study the relative fluctuations of
the link overlap with respect to the square of standard overlap. We use the parallel tempering
algorithm (PT) to investigate lattice sizes from $L=3$ to $L=12$.
For every size, we simulate at least $2048$ disorder realizations.
For the larger sizes we used $37$ temperature values in the range $0.5  \le T \le 2.3 $.
The choice of the lowest temperature is related to the possibility to thermalize the large
sizes, but our results are perfectly compatibile with those obtained by Marinari and Parisi
at $T=0$ (see the last paragraph before the conclusions).
The thermalization in the PT procedure is tested by checking the symmetry of the probability distribution
for the standard overlap $q$
under the transformation
$q \to -q$. Moreover, for the Gaussian coupling case it is available another
thermalization test: the internal energy can be calculated
both as the temporal mean of the Hamiltonian and - by exploiting integration
by parts - as expectation of a simple function of the link overlap \cite{PC}.
We checked that with our thermalization steps both measurements converge
to the same value.
All the parameters used in the simulations are reported in Tab.\ref{t:para}.
\begin{table}
\begin{center}
\begin{tabular}{|c|c|c|c|c|c|c|c|}\hline
$L$ & Therm    & Equil   & Nreal  & $n_{\beta}$ & $\delta T$ & $T_{min}$ & $T_{max}$\\ \hline \hline
$3-6$ & $50000$  & $50000$ & $2048$ &   $19$    & $0.1 $     &  $0.5$    &   $2.3$ \\ \hline
$8$   & $50000$  & $50000$ & $2680$ &   $19$    & $0.1 $     &  $0.5$    &   $2.3$ \\ \hline
$10$  & $70000$  & $70000$ & $2048$ &   $37$    & $0.05$     &  $0.5$    &   $2.3$ \\ \hline
$12$  & $70000$  & $70000$ & $2048$ &   $37$    & $0.05$     &  $0.5$    &   $2.3$ \\ \hline
\end{tabular}\caption{Parameters of the simulations: system size, number of
sweeps used for thermalization, number of sweeps for measurement of the observables, number
of disorder realizations, number of temperature values allowed in the PT procedure,
temperature increment, minimum and maximum temperature values.}\label{t:para}
\end{center}
\end{table}

We recall the basic definitions. For a 3-dimensional lattice $\Lambda$ of volume $N=L^3$,
the {\em square of the standard overlap} among two spin configurations $\s,\tau \in \{+1,-1\}^N$ is
\be
q^2(\s,\tau) \, = \, \left(\frac{1}{N}\sum_i\s_i\tau_i\right)^2.
\ee
The {\em link overlap} is instead obtained from the nearest neighbor spins, namely
for $b=(i,j)$ with $i,j\in\Lambda$, $|i-j|=1$ and $\s_b=\s_i\s_j$
\be
Q(\s,\tau) \, = \, \frac{1}{3N}\sum_b\s_b\tau_b.
\ee
First, we investigate the behavior of the correlation coefficient
between $q^2$ and $Q$
\be
\label{rho}
\rho \, = \,
\frac{<(q^2-<q^2>)(Q-<Q>)>}{\sqrt{<(q^2-<q^2>)^2> <(Q-<Q>)^2>}}.
\ee
This quantity will tell us in which range of temperatures the two random variables
are correlated. In that range we further investigate
the nature of the mutual correlation by studying their joint distribution
and, in particular, the conditional distribution $P(Q|q^2)$
of $Q$ at fixed values of $q^2$.
We are interested in understanding if a functional relation among the two quantities exists,
i.e. if the variance of the conditional distribution shrinks to zero at large volumes
and around what curve the conditional distribution is peaked.
We have:
\begin{equation}
P(Q|q^2)=\frac{P(Q,q^2)}{P(q^2)}=
\nonumber
\end{equation}
\begin{equation}
= \frac{
{\rm Av}\left(\frac{\sum_{\sigma,\tau}\delta(Q-Q_{\sigma,\tau})\delta(q^2-q^2_{\sigma,
\tau})e^{-\beta[H_\sigma+H_\tau]}}{\sum_{\sigma,\tau}e^{-\beta[H_\sigma+H_\tau]}}
\right)}
{
{\rm Av}\left(\frac{\sum_{\sigma,\tau}
\delta(q^2-q^2_{\sigma,\tau})e^{-\beta[H_\sigma+H_\tau]}}{\sum_{\sigma,\tau}
e^{-\beta[H_\sigma+H_\tau]}}
\right)}.\label{e:proQcond}
\end{equation}
For this conditional distribution one could compute the generic
$k-$th moment
\be
G_k(q^2) := \av{Q^k|q^2} = \int_{-1}^{1} Q^k P(Q|q^2) dQ.
\ee
We will be interested in the mean
\be
\label{gq2}
G(q^2):=G_1(q^2)=  \av{Q|q^2}
\ee
and the variance
\be
\label{var}
Var(Q|q^2) = G_2(q^2) - G^2_1(q^2).
\ee
The method of the least squares immediately entails that
the mean $G(q^2)$ is the best estimator for the functional
dependence of $Q$ in terms of $q^2$.
In fact, given any function  $h(q^2)$, the mean of $(Q-h(q^2))^2$
according to the joint distribution $P(Q,q^2)$  is
$\sum_{i,j}(Q_i-h(q_j^2))^2P(Q_i,q_j^2)=\sum_j
P(q_j^2)\sum_i(Q_i-h(q_j^2))^2P(Q_i|q_j^2)$, where the sums run
over all possible values of the random vector $(Q,q^2)$, which
are finitely many on the finite system we simulated. Therefore, to minimize
the mean it suffices to minimize the inner sum, i.e. to choose
$h(q^2)$ as the mean $G(q^2)$ of $Q$ with respect to the conditional
distribution (\ref{e:proQcond}).

The scaling properties of the conditional variance (\ref{var})
and the functional dependence (\ref{gq2}) provide important informations
about the low temperature phase of the model. Indeed, a vanishing
variance in the thermodynamic limit implies that the two random
variables $Q$ and $q^2$ do not fluctuate with respect to each other.
If the functional dependence $G(q^2)$ among the two is a one-to-one
increasing function, then it follows that the marginal probability
distributions for the standard and link overlap must have similar
properties. In particular, if one of the two is supported over a point
then also the other must be so.

\begin{figure}
    \setlength{\unitlength}{1cm}
    \begin{minipage}[t]{7cm}
          \centering
               \includegraphics[width=7cm,height=7cm]{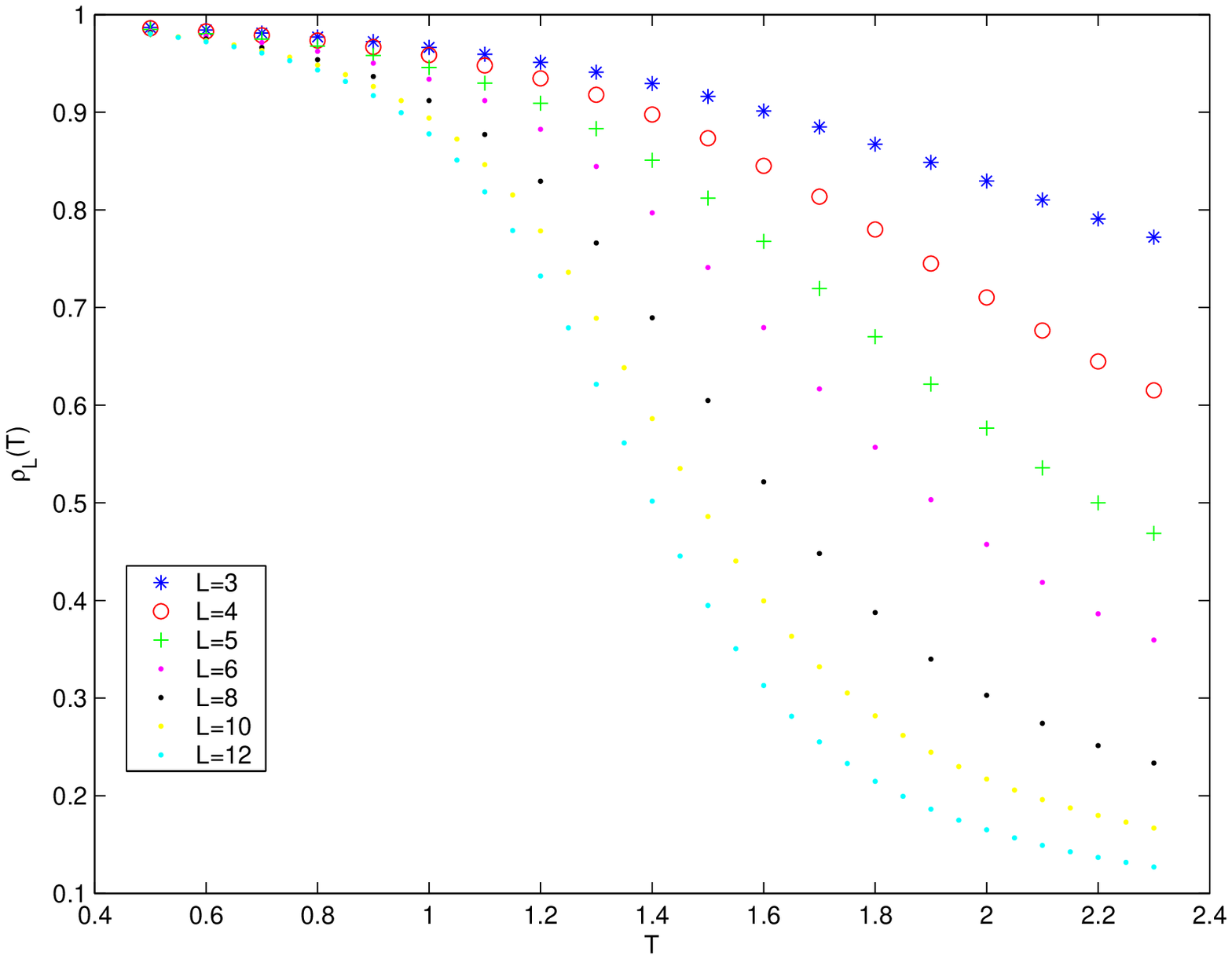}
               \caption{$\rho_L(T)$ as a function of the temperature $T$
               for different sizes $L$ of the system.}\label{fig1}
   \end{minipage}
   \ \hspace{.3cm} 
   \begin{minipage}[t]{7cm}
          \centering
               \includegraphics[width=7cm,height=7cm]{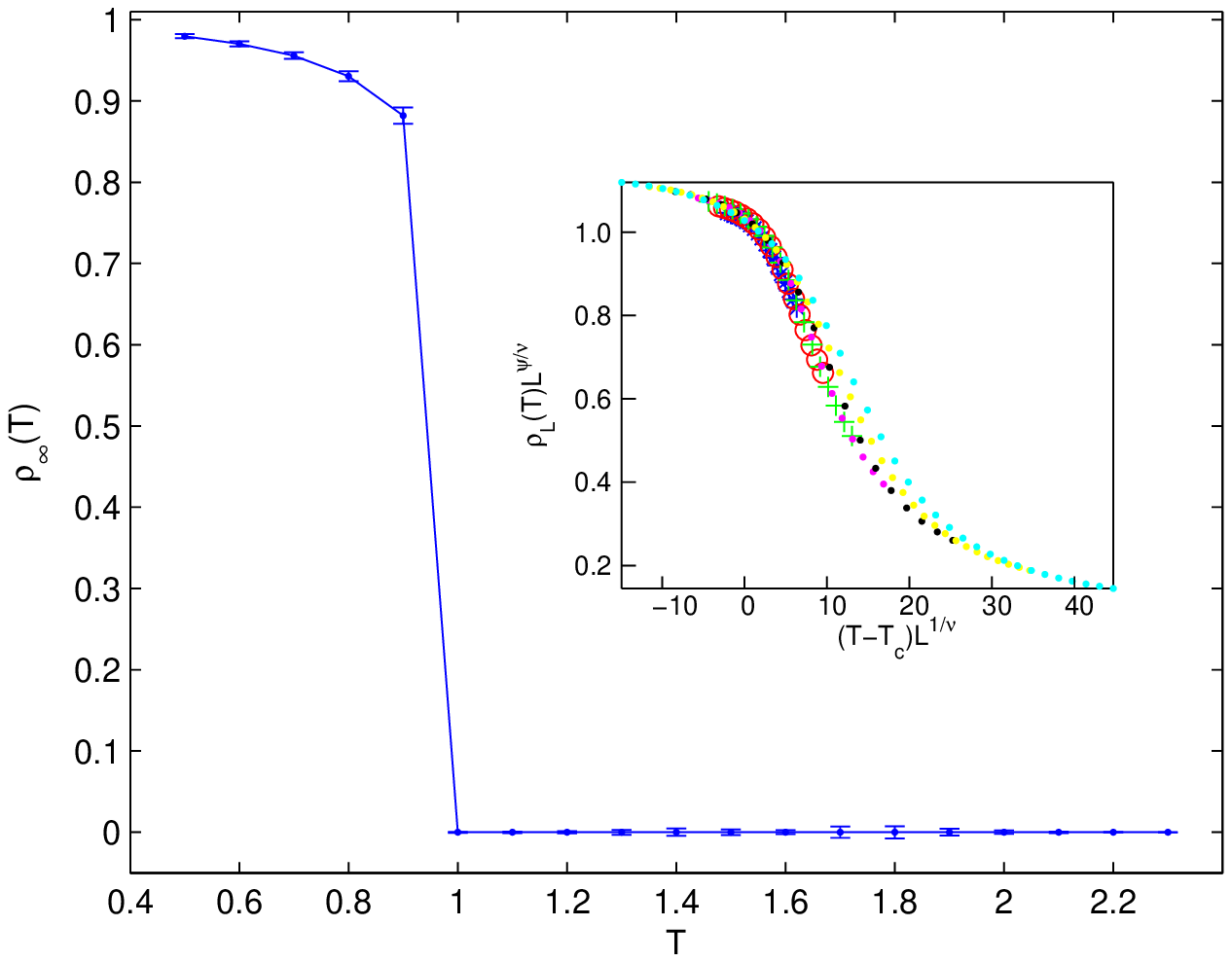}
               \caption{$\rho_{\infty}(T)$ as a function of the temperature $T$.
               In the inset it is shown the finite size scaling.}\label{fig2}
   \end{minipage}
\end{figure}

We now describe our results. Fig. (\ref{fig1}) shows the
correlation between the square standard overlap and the link
overlap. The plot of Eq. (\ref{rho}) is done for different sizes of
the system as a function of the temperature.
It is clear from the figure that, as the system size increases,
the correlation remains strong in the low-temperature region,
while it becomes weaker in the high temperature region.
A sudden change in the infinite volume behavior of $\rho$ can be expected to
occur close to the critical temperature $T_c$ of the model.
The best estimate available in the literature - obtained
through the analysis of the Binder parameter's curves of
the variable $q^2$ for different system sizes - gives
$T_c=0.95\pm 0.04$ \cite{MPRL} (we independently reproduced this
estimate with our data for $q^2$ and obtained an estimate of $T_c=0.95\pm 0.03$).

For each temperature, we did a fit of the data for $\rho$ to the infinite volume limit.
We tried different scaling for the data, both exponential
$\rho_L(T)=\rho_{\infty}(T)+a(T)e^{b(T)L}$ and power law
$\rho_L(T)= \rho_{\infty}(T) + \alpha(T)L^{\beta(T)}$. The
interesting information is contained in the asymptotic value
$\rho_{\infty}(T)$. We measured the normalized $\chi^2$ for different values
of $\rho_{\infty}(T)$ in the range $[0,\min_{L} \rho_L(T)]$ and
keeping $a(T)$ and $b(T)$ (or $\alpha(T)$ and $\beta(T)$) as free
parameters. In the region $T \geq 1.0$, we found that
$\chi^2$ attains his minimal value for $\rho_{\infty}(T) = 0$.
For $T\le 0.9$ the $\chi^2$ develops a sharp minimum corresponding
to values $\rho_{\infty}(T) \neq 0$. The whole plot of the curve
$\rho_{\infty}(T)$ as obtained from the best fit is represented in
Fig. (\ref{fig2}).
Also, in the inset of the same figure it is shown the standard finite
size scaling of the data. We plot $\rho_L(T)L^{\psi/\nu}$ versus
the scaling variable $(T - T_c)L^{1/\nu}$. A good  scaling
plot is obtained using $T_c=0.95$, $\nu=0.71$, $\psi=0.038$.
The discrepancy between the value $\nu=2.0$ of ref. \cite{MPRL}
has to be attributed to the non-linear relation between
$Q$ and $q^2$ (see below). Fig. (\ref{fig2}) tells us that
in the high temperature phase the two random variables standard and
link overlap are asymptotically
uncorrelated while in the low temperature one they display a
non-vanishing correlation: within our available discrete set of
temperature values, the temperature at which the correlation
coefficient starts to be different from zero is in good agreement
with the estimated critical value of the model.

\begin{figure}
    \setlength{\unitlength}{1cm}
    \begin{minipage}[t]{7cm}
          \centering
              \includegraphics[width=7cm,height=7cm]{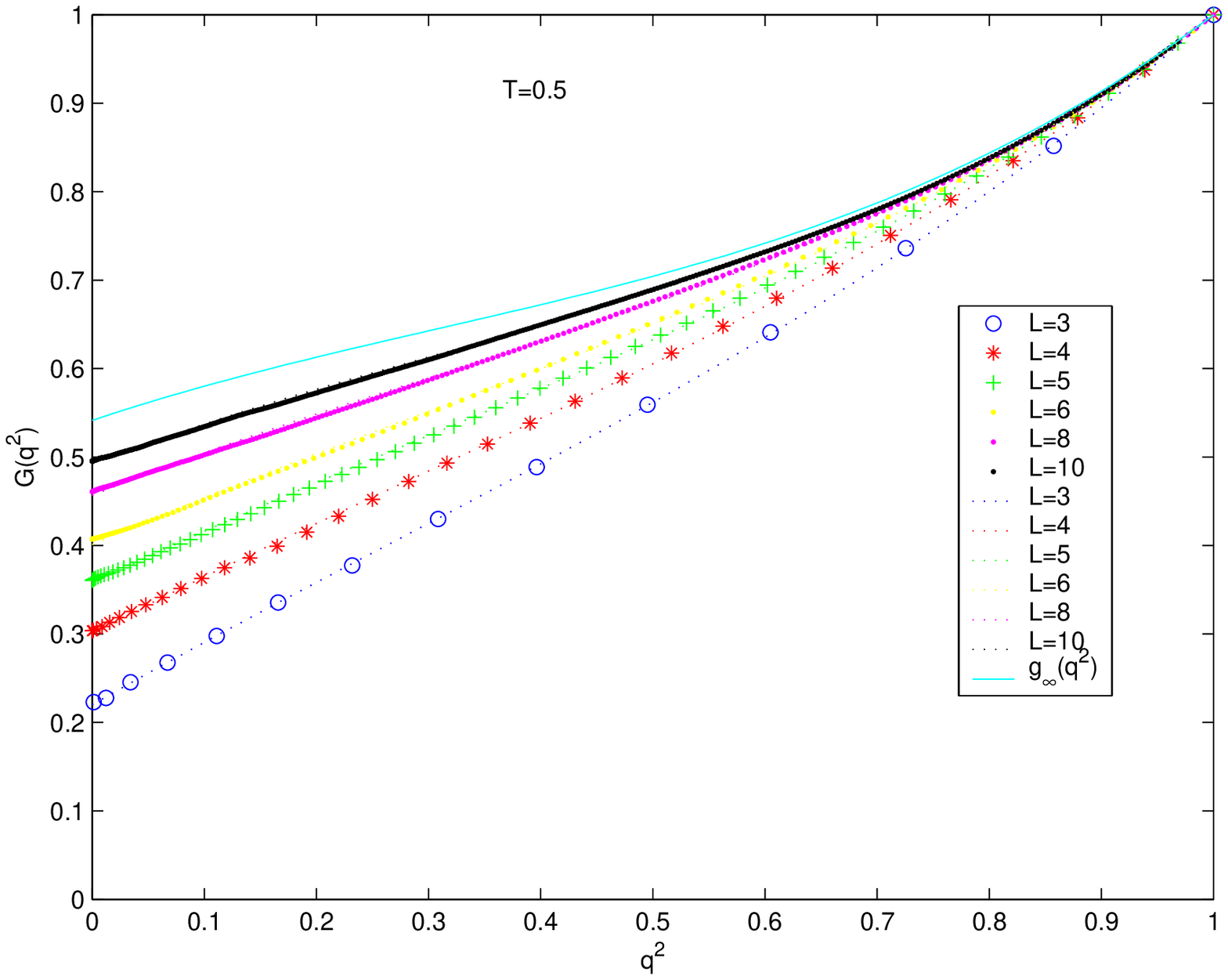}
              \caption{Plot of the curves $g(q^2)$ (continuous lines)
              and of $G(q^2)$ (dotted lines) together with the
              infinite volume limit curve $g_{\infty}(q^2)$ (upper continuous line) for $T=0.5$.}\label{fig3}
   \end{minipage}
   \ \hspace{.3cm} 
   \begin{minipage}[t]{7cm}
         \centering
               \includegraphics[width=7.2cm,height=7.2cm]{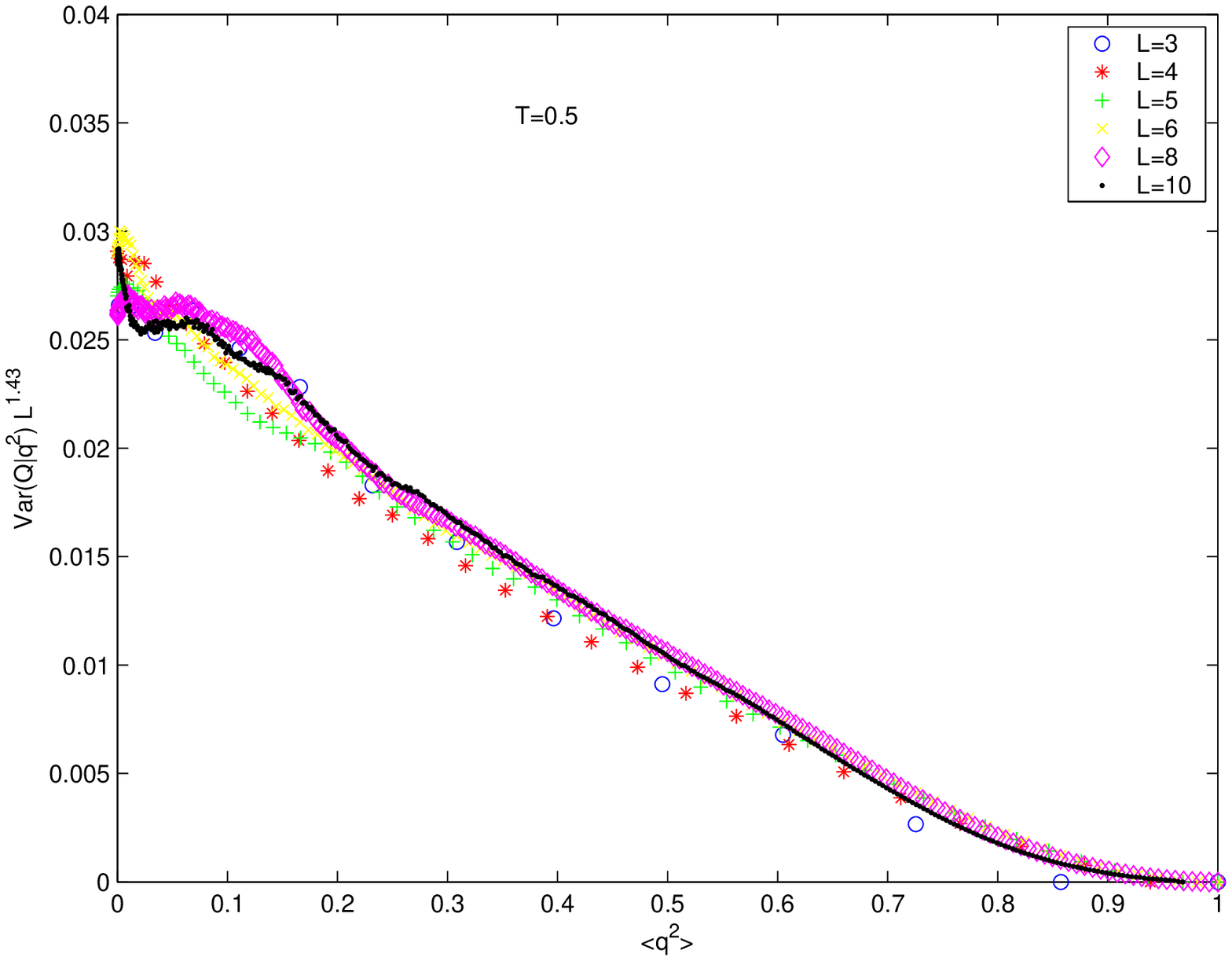}
               \caption{$L^{1.43}Var_L(Q|q^2)$ as a function of $q^2$ for the temperature $T=0.5$
               and for different sizes $L$.}\label{fig4}
   \end{minipage}
\end{figure}

We consider then the problem of studying the functional dependence
(if any) between the two random variables $Q$ and $q^2$ in the low
temperature region.
The points in the Fig. (\ref{fig3}) show the function $G(q^2)$ of Eq.(\ref{gq2})
for different system sizes at $T=0.5$, well below the critical temperature.
Also we studied a third order approximation of the form
$Q=g(q^2)=a+bq^2+cq^4+dq^6$. Since we must have  $Q=1$
for $q^2=1$, this actually implies $d=1-a-b-c$.
The coefficients $a_{L,T},b_{L,T},c_{L,T}$
have been obtained by the least square method and then fitted
to the infinite volume limit.
The result is shown as continuous lines
in Fig (\ref{fig3}). The good superposition of the curves to the data for
$G(q^2)$ indicates that the functional dependence between the
two overlaps is well approximated already at the third order.

Finally, we measured the variance Eq. (\ref{var}) at low temperatures.
We observed that the distribution is concentrating for large volumes around its mean value.
The trend toward a vanishing variance for infinite system
sizes is very clear. We analyzed {\em all} temperatures below $T_c$
and we found that the best fit of $Var_L(Q|q^2)$, in terms of the $\chi^2$,
is obtained by a power law of the form
$a(T)L^{-b(T)}+c$ and it gives $c=0$ for every value of the temperature.
For the lowest available temperature $T=0.5$, this is shown in Fig (\ref{fig4})
where we plot the data for $Var_L(Q|q^2) L^{1.43}$ for different system sizes $L$:
all the different curves collapse to a single one.
The data for other temperature values behave similarly,
the only difference being that the coefficient $b(T)$ is increasing
with the temperature (it stays in the range $[1.43,1.74]$ for $T\in[0.5,0.9]$).
This result has quite strong consequences because it
says that the two random variables $Q$ and $q^2$ cannot have different triviality
properties: if one of them is trivial (delta-like distributed) the scaling law for their
conditional variance implies that also the other is trivial. Our result rules then
out the possibility to have a non-trivial standard overlap with a trivial link overlap
as predicted for instance in the so called TNT picture \cite{KM,PY}.

It is interesting to compare our result with previous works.
Marinari and Parisi \cite{MP} have studied the relation
$Q=(1-A(L))+(A(L)-B(L))q^2+B(L)q^4$
among the two
overlaps {\em at zero temperature}, by ground state perturbation.
We have extrapolated our data in the
low temperature regime to zero temperature by a polynomial fit
and then to the infinite volume limit ($L=\infty$).
The best fit for $L=\infty$ (i.e. the one with smaller $\chi^2$)
is quadratic in $L^{-1}$. It gives $A=0.30\pm 0.05$ ($\chi^2=0.21$),
which is in agreement with the independent measure of Marinari
and Parisi ($A=0.30\pm 0.01, \chi^2=0.6$).
Note that their results are obtained with a complete different
method than Montecarlo simulations, namely exact ground states.
Sourlas \cite{Sou} studied the same problem in a different setting
called soft constraint model. Although a direct quantitative comparison
is not possible with our method, his results are qualitatively similar.
In the context of out-of-equilibrium dynamics, a strong correlation
between link-overlap and standard-overlap in the low-temperature
phase was pointed out in ref. \cite{JMPT}.

In conclusion, our result shows quite clearly that, within the tested system sizes,
the study of the two order parameters, the square of the standard overlap and the link overlap,
are equivalent as far as the quenched equilibrium state is concerned.
In view of our result, the proposed pictures which assign different
behaviour to the two overlap distributions, in particular the TNT \cite{KM,PY} picture, should
be rejected. It is interesting to point out that, since the present analysis
deals only with the distribution of $P(q^2,Q)$ and not with the higher order ones
like for instance $P(q^2_{1,2},q^2_{2,3},Q_{1,2},Q_{2,3})$, our results are compatible
with different factorization properties of the two overlaps
like those illustrated in \cite{CG}.\\

{\bf Acknowledgments.} We thank Enzo Marinari, Giorgio Parisi and Federico Ricci Tersenghi
for interesting discussions. We thank a referee for the suggestion to improve
the Fig. (\ref{fig2}) by a finite scale analysis. We also thank Cineca for the grant in computer time.
C. Giardin\`a acknowledges NWO-project 613000435 for financial support.

\newpage

\begin{thebibliography}{99}

\bibitem{MPV}
M. Mezard, G. Parisi, M.A. Virasoro,
{\em Spin Glass Theory and Beyond}
World Scientific, Singapore (1987).

\bibitem{NS} C.M. Newman and D.L. Stein,
{\it cond-mat/0503345}.

\bibitem{EA} S.F. Edwards and P.W.Anderson,
{\em Jou. Phys. F.}, {\bf 5}, 965, (1975).

\bibitem{Pa} G. Parisi, F. Ricci-Tersenghi,
{\em J. Phys. A: Math. Gen.} {\bf 33}, 113, (2000).

\bibitem{PC} P. Contucci,
{\em J. Phys. A: Math. Gen.} {\bf 36}, 10961, (2003).



\bibitem{MP} E. Marinari, G. Parisi,
{\em Phys. Rev. Lett.} {\bf 86}, 3887, (2001).

\bibitem{MPRL} E. Marinari, G. Parisi, and J.J. Ruiz-Lorenzo,
{\em Phys. Rev. B} {\bf 58}, 14852 (1998)

\bibitem{JMPT}
S. Jimenez, V. Martin-Mayor, G. Parisi, A. Tarancon
{\em J. Phys. A: Math. and Gen.} {\bf 36}, 10755 (2003)

\bibitem{Sou} N. Sourlas,
{\em Phys. Rev. Lett.} {\bf 94}, 70601, (2005).

\bibitem{KM}
F. Krzakala and O. C. Martin,
{\em Phys. Rev. Lett.} {\bf 85}, 3013 (2000)

\bibitem{PY} M.Palassini, A.P.Young,
{\em Phys, Rev. Lett.} {\bf 85}, 3017, (2000).

\bibitem{FH} D.S. Fisher and D.A. Huse,
{\em Phys. Rev. Lett.} {\bf 56}, 1601 (1986)

\bibitem{CG} P.Contucci, C.Giardin\`a,
{\em Phys. Rev. B}  {\bf 72}, 14456, (2005).

\end{thebibliography}
\end{document}